\documentclass{ifacconf}

\usepackage{graphicx}      
\usepackage{wrapfig}
\usepackage{natbib}        
\usepackage{amsmath} 
\usepackage{amssymb}  
\usepackage{amsfonts}
\usepackage{tikz}
\usepackage{caption}
\usepackage{subcaption}
\usepackage{color}
\usepackage{comment}

\newcommand{\dfb}{\stackrel{\Delta}{=}}
\newcommand{\R}{\ensuremath{\mathbb R}}

\newtheorem{remark}{Remark}

\begin{document}
\begin{frontmatter}

\title{Controlling a triangular flexible formation of autonomous agents.\thanksref{footnoteinfo}} 

	\thanks[footnoteinfo]{The work of Hector Garcia de Marina was supported by Mistrale project, http://mistrale.eu. The work of Zhiyong Sun and Brian Anderson was supported by the Australian Research Council Grants DP160104500, DP130103610 and the Prime Minister’s Australia Asia Incoming Endeavour Postgraduate Award. The work of Cao was supported in part by the European Research Council (ERC-StG-307207) and the Netherlands Organization for Scientific Research (NWO-vidi-14134). E-mails: hgdemarina@ieee.org; \{zhiyong.sun,brian.anderson\}@anu.edu.au; m.cao@rug.nl.}

\author[First]{Hector Garcia de Marina}
\author[Second]{Zhiyong Sun}
\author[Third]{Ming Cao}
\author[Forth,Second]{Brian D.O. Anderson}

\address[First]{University of Toulouse, Ecole Nationale de l'Aviation Civile (ENAC) Toulouse 31000, France.}
\address[Second]{Data61-CSIRO and Research School of Engineering, Australian National University, Canberra ACT 2601, Australia.}
\address[Third]{Engineering and Technology Institute Groningen (ENTEG), University of Groningen 9747, the Netherlands.}
\address[Forth]{School of Automation, Hangzhou Dianzi University, Hangzhou
310018, China.}

\begin{abstract}                
	In formation control, triangular formations consisting of three autonomous agents serve as a class of benchmarks that can be used to test and compare the performances of different controllers. We present an algorithm that combines the advantages of both position- and distance-based gradient descent control laws. For example, only two pairs of neighboring agents need to be controlled, agents can work in their own local frame of coordinates and the orientation of the formation with respect to a global frame of coordinates is not prescribed. We first present a novel technique based on adding artificial biases to neighboring agents' range sensors such that their eventual positions correspond to a collinear configuration. Right after, a small modification in the bias terms by introducing a prescribed rotation matrix will allow the control of the bearing of the neighboring agents.
\end{abstract}

\begin{keyword}
Formation control, Distributed control, Multi-agent system.
\end{keyword}

\end{frontmatter}

\section{Introduction}
The theory of rigidity and in particular the concepts of \emph{infinitesimal and minimal rigidity} have been proven to be very useful in formation control with the goal to define and achieve prescribed shapes by just controlling the distances between neighboring agents (see \cite{AnYuFiHe08} and \cite{KrBrFr08}). The popular distance-based gradient descent algorithm for rigid formations has appealing properties. For example, the agents can work in their own local frame of coordinates, the system can be made robust against biased sensors, and since the orientation of the desired shape is not prescribed, one can induce rotational motions to the formation as presented in \cite{MarCaoJa15, Hector2016maneuvering}. In comparison, position-based formation control needs a common frame of coordinates and the steady-state orientation of the shape is restricted, which implies that a free rotational motion is not allowed; however, the same formation shape can be achieved by fewer pairs of neighboring agents in position-based than in distance-based control. This prompts a search for compromise among the advantages of both formation control techniques.

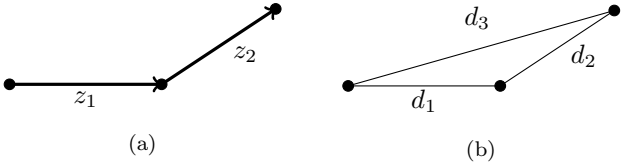
\begin{figure}
\centering

\begin{subfigure}{0.49\columnwidth}
\centering
\begin{tikzpicture}[line join=round]
\filldraw(0,0) circle (2pt);
\filldraw(2,0) circle (2pt);
\filldraw(3.5,1) circle (2pt);
\draw[very thick,arrows=->](0,0)--(2,0);
\draw[very thick,arrows=->](2,0)--(3.5,1);
\node at (1,-.2) {$z_1$};\node at (3.1,.4) {$z_2$};
\end{tikzpicture}
\caption{}
\end{subfigure}
\begin{subfigure}{0.49\columnwidth}
\centering
\begin{tikzpicture}[line join=round]
\filldraw(0,0) circle (2pt);
\filldraw(2,0) circle (2pt);
\filldraw(3.5,1) circle (2pt);
\draw[](0,0)--(2,0);
\draw[](2,0)--(3.5,1);
\draw[](3.5,1)--(0,0);
\node at (1,-.2) {$d_1$};\node at (3.1,.4) {$d_2$};\node at (1.7,.9) {$d_3$};
\end{tikzpicture}
\caption{}
\end{subfigure}

	\caption{The same triangular shape depicted by the three agents (dots) can be achieved by controlling the two vectors $z_1$ and $z_2$ (position-based control) or the three distances $d_1$, $d_2$ and $d_3$ (distance-based control). Note that in a) the orientation of the triangle is prescribed while it is not in b).}
\label{fig: triangulos}
\end{figure}

The minimum total number of pairs of neighboring agents in a distance-based rigid formation control system is given by the necessary conditions for infinitesimal and minimal rigidity, which is equal to $2n - 3$ for 2D scenarios, where $n$ is the total number of agents. This is directly related to the necessary number of inter-agent distances to be controlled such that the desired shape is (at least locally) uniquely defined. For example, the triangle is the simplest rigid shape in the 2D case and the necessary number of desired distances to define such a shape is three. On the other hand, one can achieve a desired triangular shape by just controlling (the relative positions rather than distances of) two neighboring agent pairs in the position-based setup. We illustrate these basic concepts in Figure \ref{fig: triangulos}.

This paper focuses on triangular formations consisting of three agents. This apparently simple setup has been considered as a benchmark in formation control (see \cite{cao2007controlling, anderson2007control, cao2008generalized, liu2014controlling, 7039453}), since it allows detailed rigorous analysis for novel techniques as we aim at in this work. This provides a starting point in order to achieve more general formations. The goal of this paper is to propose an algorithm that combines the advantages of both distance- and position-based control, i.e., agents employ their own local frames of coordinates, no prescribed orientation for the desired shape (so rotational motions are allowed) and smaller number of controlled pairs of neighboring agents than in distance-based minimally rigid formations.

The novelty of this algorithm lies in exploiting the sensing-error-induced collective motion, e.g., those caused by biased range sensors among neighboring agents. For example, it has been reported in \cite{SMA16TACsub} that small constant biases in range sensors for a rigid formation cause the whole team of agents to converge to a distorted version of the desired shape and eventually it will exhibit some steady-state motion. On the other hand, it has been shown in \cite{Hector2016maneuvering} that one can introduce artificially such biases in order to steer the desired rigid formation in a controlled way. 

This paper will first study the effect of these biases in a \emph{flexible} (or non-rigid) distance-based formation of three agents, i.e., we control only two distances instead of three for a rigid triangular formation. We prove that if the range measurements between agents are not perfectly accurate then the three agents converge to collinear positions. This is somewhat counter-intuitive since for a non-rigid formation the steady-state shape is, depending on the initial conditions, arbitrary within a constraint set. Furthermore, we will show that depending on the value of these biases, the eventual collinear formation will move with a constant velocity or remain stationary.
 In our first step, we take advantage of this effect in order to align the three agents with the desired distances between them. Then, we will show how a small modification in this \emph{biased} control law can achieve the control of the angle between the two relative vectors of the flexible formation, leading to the control of a stationary desired triangular shape with no restriction on its orientation.

In the following section we introduce some notations, briefly review the (exponential) stability of formation systems under the \emph{flexible} shape and introduce the biases in the range measurements. We continue in Section \ref{sec: biased} analyzing the consequences of such biases in the control of a non-rigid formation. We introduce a technique motivated by \cite{SMA16TACsub}, where the authors study non-minimally but infinitesimally rigid formations by \emph{reducing} the problem to minimally rigid ones. In this paper since we are dealing with fewer pairs of neighboring agents than in minimally rigid setups, we instead \emph{augment} our formation to a minimally rigid one in order to study the stability of a biased non-rigid formation. In Section \ref{sec: trian} we introduce and analyze a modified version of the \emph{biased} algorithm in order to control triangular shapes. We finish the paper with simulations in Section \ref{sec: sim} and some conclusions in Section \ref{sec: con}.

\section{Non-rigid formation of three agents}

\label{sec: empezamos}

\subsection{Gradient descent distance-based formation control}
We consider a team of three agents governed by the first-order kinematic model
\begin{equation}
	\dot p_i = u_i,
	\label{eq: pdyn1}
\end{equation}
where $p_i\in\mathbb{R}^2$ is the position of the agent $i=\{1,2,3\}$, and $u_i\in\mathbb{R}^2$ is the control action over agent $i$.

Let us define the two vectors
\begin{equation}
z_1 \dfb p_1 - p_2, \quad z_2 \dfb p_2 - p_3,
\end{equation}
which are the relative positions corresponding to the two links available to the agents. For each link one can construct a potential function $V_k, k\in\{1,2\}$ with its minimum at the desired distance $d_k$, so that the gradient of such functions can be used to control inter-agent distances distributively. We consider the following \emph{shape potential function}
\begin{equation}
	V_k(z_k) = \frac{1}{2}(||z_k|| - d_k)^2,
	\label{eq: Vkquad}
\end{equation}
with the following gradient along $z_k$
\begin{equation}
	\nabla_{z_k}V_k(z_k) = \hat z_k(||z_k|| - d_k),
	\label{eq: zgrad}
\end{equation}
where $\hat z_k \dfb \frac{z_k}{||z_k||}$. We then apply to each agent $i$ in (\ref{eq: pdyn1}) the following gradient descent control
\begin{equation}
	u_i = -\nabla_{p_i}\sum_{k=1}^{2} V_k(z_k), 
	\label{eq: udyn1}
\end{equation}
and by denoting the distance error for the $k$th link by
\begin{equation}
	e_k = ||z_k|| - d_k,
	\label{eq: ek}
\end{equation}
we arrive at the following dynamics
\begin{equation}
	\begin{cases}
	\dot p_1 &= -\hat z_1 e_1 \\
	\dot p_2 &= \hat z_1 e_1 -\hat z_2 e_2 \\
	\dot p_3 &= \hat z_2 e_2.
	\end{cases}
	\label{eq: 3flex}
\end{equation}
System (\ref{eq: 3flex}) has some interesting properties. For example, the agents can employ their own local systems of coordinates, i.e., a global or common frame of coordinates is not necessary, and collision avoidance is guaranteed among pairs of neighboring agents (\cite{oh2015survey}). For \emph{tree graph} topology, one can prove the (almost global) exponential convergence of the error system, i.e., the signals $e_1(t)$ and $e_2(t)$ both converge to zero exponentially fast if the agents do not start at the same positions (\cite{dimarogonas2008stability,sun2016exponential}).

It is clear that although the error signals $e_1(t)$ and $e_2(t)$ converge to zero, the final relative positions may not guarantee any prescribed shapes as one would like to have for minimally rigid formations. In particular, agents $p_1$ and $p_3$ will lie somewhere on a circumference with the center at $p_2$ and the radius as $d_1$ and $d_2$ respectively. More precisely, the relative positions converge to the set 
\begin{equation}
\mathcal{Z} \dfb \{z : ||z_k|| = d_k, \forall k\in\{1,2\} \}.
	\label{eq: noshape}
\end{equation}
Furthermore, the exponential convergence of $e(t)$ and $z(t)$ to the origin and $\mathcal{Z}$, respectively, implies that $\dot p_1(t)$ and $\dot p_2(t)$ converge to zero exponentially fast, therefore the agents converge to only one stationary shape among all the ones defined by $\mathcal{Z}$, i.e., all the agents eventually stop.

\subsection{Biases in distance-based formation control}

It is clear from (\ref{eq: udyn1}) with $V_k$ as in (\ref{eq: Vkquad}) that neighboring agents $i$ and $j$ share the same potential $V_k$ in the implementation of the gradient formation control. Let us focus on agent $i$ whose control is
\begin{equation*}
	u_i = -\sum_{k=1}^{2} u^k_i,
\end{equation*}
where $u^k_i$ is the corresponding gradient $\nabla_{p_i} V_k$. More precisely, for the link $k$ where $i$ and $j$ are neighbors we have that
\begin{align}
	u^k_i &= \nabla_{p_i}V_k = \hat z_k (||z_k|| - d_k) \label{eq: uki} \\
	u^k_j &= \nabla_{p_j}V_k = -\hat z_k (||z_k|| - d_k) \label{eq: ukj}.
\end{align}
However, when the range sensors of the two agents are biased with respect to each other by a constant $\mu_k$, without loss of generality, the pair (\ref{eq: uki})-(\ref{eq: ukj}) should be modified as
\begin{align}
	u^k_i &= \hat z_k (||z_k|| - d_k + \mu_k) \label{eq: ukib} \\
	u^k_j &=  -\hat z_k (||z_k|| - d_k) \label{eq: ukjb}.
\end{align}
The effects of these biases on undirected rigid formations have been studied in \cite{SMA16TACsub} and how to remove or take advantage of them have been shown in \cite{MarCaoJa15, Hector2016maneuvering}. Flexible formations have not received much attention for formation control problems as the rigid ones since the sets like (\ref{eq: noshape}) do not define (locally) uniquely any shape. Nevertheless, we will show that the intentional introduction of biases (following the spirit of \cite{Hector2016maneuvering}) can make these apparently flexible formations appealing for formation control.

\section{Biased non-rigid formation of three agents: converging to a collinear steady-state}
\label{sec: biased}
Let us consider biases in the range sensors of the agents in system (\ref{eq: 3flex}). In particular, as it has been shown in (\ref{eq: ukib})-(\ref{eq: ukjb}), we can always write these biases from the point of view of agent $2$, namely
\begin{equation}
	\dot p_2 = \hat z_1 (e_1+\mu_1) -\hat z_2 (e_2-\mu_2),
\end{equation}
which leads to the following \emph{biased} system
\begin{equation}
	\begin{cases}
	\dot p_1 &= -\hat z_1 e_1 \\
	\dot p_2 &= \hat z_1 e_1 -\hat z_2 e_2 + \mu_1\hat z_1 + \mu_2 \hat z_2\\
	\dot p_3 &= \hat z_2 e_2.
	\end{cases}
	\label{eq: 3flexmu}
\end{equation}
We shall now study the effect of imposing a special condition on the biases. This condition will be achievable if we are free to introduce biases into the controls, namely
\begin{equation}
\mu_1 = -\mu_2 = c,
	\label{eq: muc}
\end{equation}
for some constant $c \in\R\backslash \{0\}$. These biases can be considered as a parametric disturbance to the following closed-loop system involving the dynamics of $z$ and $e$ derived from (\ref{eq: 3flexmu}):
\begin{equation}
	\begin{cases}
		\dot z_1 &= -2\hat z_1e_1 + \hat z_2e_2 - c\hat z_1 + c\hat z_2 \\
		\dot z_2 &= -2\hat z_2e_2  + \hat z_1e_1 + c\hat z_1 - c\hat z_2 \\
		\dot e_1 &= -2e_1 + \hat z_1^T\hat z_2e_2 - c + c\hat z_1^T \hat z_2 \\
		\dot e_2 &= -2e_2 + \hat z_2^T\hat z_1e_1 - c + c\hat z_2^T \hat z_1.
	\end{cases}
	\label{eq: ezdynm}
\end{equation}
As we have discussed, one can derive the exponential stability of the closed-loop system (\ref{eq: ezdynm}) with $c = 0$ (\cite{dimarogonas2008stability}). Therefore the stability of (\ref{eq: ezdynm}) for small values of $c$ is not compromised but at the cost of (probably) shifting its equilibrium from the one described in (\ref{eq: noshape}).

One realizes by a quick inspection in (\ref{eq: 3flexmu}) together with (\ref{eq: muc})\footnote{For arbitrary values of $\mu_1$ and $\mu_2$ the results of this paper apply as well but with different steady-state values for $e_1$ and $e_2$. For the sake of simplicity and clarity we only consider the case where (\ref{eq: muc}) holds. For more details, we refer to \cite{SMA16TACsub} and \cite{Hector2016maneuvering}.} that $e_1 = e_2 = 0$ with $\hat z_1 = \hat z_2 = \hat z^*$ is an equilibrium for the system (\ref{eq: ezdynm}), implying that all the agents are stationary. Note that the agents are collinear with agent $2$ in the middle for this configuration. We define this equilibrium for (\ref{eq: ezdynm}) by
\begin{equation}
	\mathcal{U}_d \dfb \{z, e \, : \, (e_1 = e_2 = 0) \wedge (\hat z_1 = \hat z_2) \}.
	\label{eq: ud}
\end{equation}
However, a more careful inspection of (\ref{eq: ezdynm}) reveals that there is another equilibrium corresponding to $e_1 = e_2 = -\frac{2c}{3}$ with $\hat z_1 = -\hat z_2 = \hat z^*$, i.e., agents $1$ and $3$ converge to the same side of agent $2$. This implies by inspecting (\ref{eq: 3flexmu}) that the whole formation is travelling with velocity $\dot p_i = \frac{2c}{3}\hat z^*, \forall i=\{1, 2, 3\}$. We define this equilibrium for (\ref{eq: ezdynm}) by
\begin{equation}
	\mathcal{U}_u \dfb \{z, e \, : \, (e_1 = e_2 = -\frac{2c}{3}) \wedge (\hat z_1 = -\hat z_2) \}.
\end{equation}

\begin{prop}
Consider the closed-loop system (\ref{eq: ezdynm}) derived from the biased flexible formation control (\ref{eq: 3flexmu}). If $\mu_1 = -\mu_2 = c$ with $c\in\R \backslash \{0\}$, then the only two equilibrium sets for (\ref{eq: ezdynm}) are $\mathcal{U}_d$ and $\mathcal{U}_u$.
	\label{pro: equi}
\end{prop}
\begin{pf}
By using the last two equations of (\ref{eq: ezdynm}) one has that the following condition holds at the equilibrium
\begin{equation}
	(2 + \hat z_1^T\hat z_2)(e_2 - e_1) = 0,
	\label{eq: cond1}
\end{equation}
from which one can derive that regardless of $\hat z_1$ and $\hat z_2$, one has that $e_1 = e_2 = e^*$ at an equilibrium. Note that this condition implies that the norms of $z_1$ and $z_2$ at the equilibrium are constant. We are going to check that such a situation can only occur at the equilibrium. For example, it might be possible that both vectors could be rotating while keeping their norms constant. Let us write the first two equations of system (\ref{eq: ezdynm}) and require that the error signals are constant
\begin{equation}
	\frac{\mathrm{d}}{\mathrm{dt}}
	\begin{bmatrix}
	z_1 \\ z_2
	\end{bmatrix} =
	\left(
	\begin{bmatrix}
		-\frac{2e^*-c}{||z_1||^*} & \frac{e^*+c}{{||z_2||^*} } \\
		\frac{e^* + c}{||z_1||^*} & -\frac{2e^*-c}{||z_2||^*} 
	\end{bmatrix}
	\otimes I_2 \right)
	\begin{bmatrix}
	z_1 \\ z_2
	\end{bmatrix},
	\label{eq: zdynM}
\end{equation}
	where $I_2$ is the $2\times 2$ identity matrix, the symbol $\otimes$ denotes the Kronecker product, and we have that $||z_1||^*$ and $||z_2||^*$ are constant since $e_1 = e_2 = e^*$. Note that we have split $\hat z_{\{1,2\}} = \frac{1}{||z_{\{1,2\}}||^*} z_{\{1,2\}}$.
	It is clear that the matrix in (\ref{eq: zdynM}) cannot be skew-symmetric since the off-diagonal terms always have the same sign, therefore we discard the possibility of rotations for $z_1$ and $z_2$ when the error signals are constant. Therefore $\dot e_1 = \dot e_2 = 0$ implies that the time derivatives $\dot z_1 = \dot z_2 = 0$ as well. By using the first two equations in (\ref{eq: ezdynm}) we can derive the following condition at an equilibrium of (\ref{eq: ezdynm})
\begin{equation}
	(\hat z_2^* - \hat z_1^*)(3e^* + 2c) = 0,
	\label{eq: zeq}
\end{equation}
where $z_1^*$ and $z_2^*$ are two fixed vectors for $z_1$ and $z_2$ such that $e_1^* = e_2^* = e^*$.
	
It is clear from (\ref{eq: zeq}) that we only have two possible equilibrium sets. Let us check the case $\hat z_1^* = \hat z_2^*$. By checking the equilibrium of the third equation in (\ref{eq: ezdynm}) one can quickly derive that $e^* = 0$. So we have checked the equilibrium $\mathcal{U}_d$. For the second case $e^* = -\frac{2c}{3}$, one can check in the first equation in (\ref{eq: ezdynm}) that for its equilibrium the relation $\hat z_1^* = -\hat z_2^*$ must be satisfied. So we have checked the equilibrium $\mathcal{U}_u$. {$\blacksquare$}
\end{pf}

One may wonder about the stability of $\mathcal{U}_d$ and $\mathcal{U}_u$. In fact, if one is interested in having the three agents in a collinear fashion while they are stationary it would be desirable to know that at least $\mathcal{U}_d$ is locally stable. We will see this fact in more detail in the following subsection.

\subsection{Stability analysis of the error system}
By inspecting the system (\ref{eq: ezdynm}) one realizes that the error system for the flexible formation is not autonomous. This is due to the fact that the term
\begin{equation}
	\hat z_1^T\hat z_2 = \frac{1}{2||z_1||||z_2||}(||p_3-p_1||^2 - ||z_1||^2 - ||z_2||^2), \label{eq: zz}
\end{equation}
depends on $||p_3-p_1||^2$ and it cannot be written as a function of $e_1$ and $e_2$. It would be desirable to handle a system involving only scalar variables than a mixture of scalar and vectorial ones as in (\ref{eq: ezdynm}). The key idea for starting the stability analysis of (\ref{eq: ezdynm}) is by including the following states
\begin{equation}
	e_3 \dfb ||z_3|| - d_3, \quad z_3 \dfb p_3 - p_1, \nonumber
\end{equation}
where one can choose the value for $d_3$ depending on the equilibrium to be studied. For example, for the equilibrium $\mathcal{U}_d$ we set $d_3 = d_1 + d_2$. The inclusion of these states will make the \emph{augmented} system for the error vector $e \dfb \begin{bmatrix}e_1 & e_2 & e_3\end{bmatrix}^T$ autonomous and therefore easier to analyze.

Let us first write the dynamics of $e$ derived from (\ref{eq: 3flexmu})
\begin{equation}
	\begin{cases}
		\dot e_1 &= -2e_1 + \hat z_1^T\hat z_2e_2 - c + c\hat z_1^T \hat z_2 \\
		\dot e_2 &= -2e_2 + \hat z_2^T\hat z_1e_1 - c + c\hat z_2^T \hat z_1 \\
		\dot e_3 &= \hat z_3^T\hat z_1e_1 + \hat z_3^T\hat z_2e_2.
	\end{cases}
	\label{eq: edynaug}
\end{equation}

By employing the relation (\ref{eq: zz}) and similarly for $\hat z_3^T\hat z_2$ and $\hat z_3^T\hat z_1$ one can derive the following self-contained dynamics for the error system
\begin{equation}
	\begin{cases}
		\dot e_1 &= -2e_1 - c \\
		&+ \frac{(e_2 + c)\left((e_3+d_3)^2 
		- (e_1 + d_1)^2 - (e_2 + d_2)^2 \right)}{2(e_1+d_1)(e_2+d_2)} \\
		\dot e_2 &= -2e_2 - c \\ 
		&+ \frac{(e_1 + c)\left((e_3+d_3)^2 
		- (e_1 + d_1)^2 - (e_2 + d_2)^2 \right)}{2(e_1+d_1)(e_2+d_2)}\\
	\dot e_3 &= \frac{e_1\left((e_2+d_2)^2 
		- (e_1 + d_1)^2 - (e_3 + d_3)^2 \right)}{2(e_1+d_1)(e_3+d_3)} \\ &+ \frac{e_2\left((e_1+d_1)^2 - (e_2 + d_2)^2 - (e_3 + d_3)^2 \right)}{2(e_2+d_2)(e_3+d_3)}.
	\end{cases}
	\label{eq: eaug}
\end{equation}
Now we linearize the system (\ref{eq: eaug}) at the equilibrium $\mathcal{U}_d$ or equivalently $e = 0$ for $d_3 = d_1 + d_2$. We first compute the partial derivatives of
\begin{align}
	f(e) &= \frac{(e_3+d_3)^2 -(e_1+d_1)^2-(e_2+d_2)^2}{2(e_1+d_1)(e_2+d_2)} \\
	\frac{\partial f(e)}{\partial{e_1}}|_{e = 0} &= 
	-\frac{\left((e_3+d_3)^2-(e_1+d_1)^2-(e_2+d_2)^2\right)}{2(e_1+d_1)^2
	(e_2+d_2)}|_{e=0} \nonumber \\
	&- \frac{2(e_1+d_1)}{2(e_1+d_1)(e_2+d_2)}|_{e=0}
	\nonumber \\ &= -\frac{d_3^2-d_1^2-d_2^2}{2d_1^2d_2}-\frac{1}{d_2} = -\frac{d_1+d_2}{d_1d_2} \nonumber \\
	\frac{\partial f(e)}{\partial{e_2}}|_{e = 0} &= -\frac{d_3^2-d_1^2-d_2^2}{2d_1d_2^2}-\frac{1}{d_1} = -\frac{d_1+d_2}{d_1d_2} \nonumber \\
	\frac{\partial f(e)}{\partial{e_3}}|_{e = 0} &= \frac{d_3}{d_1d_2}=
	\frac{d_1+d_2}{d_1d_2}, \nonumber
\end{align}
that in combination with
\begin{equation}
	f(0) = \frac{d_3^2 - d_1^2 - d_2^2}{2d_1d_2} = 1,
\end{equation}
helps us to arrive at the Jacobian matrix defining the linearization of the autonomous system (\ref{eq: eaug}) at $e=0$ as below
\begin{equation}
	J_e = \begin{bmatrix}
	-2-ca & 1-ca & ca \\
	1-ca & -2-ca & ca \\
	-1 & -1 & 0
	\end{bmatrix},\label{eq: Je}
\end{equation}
where $a = \frac{d_1+d_2}{d_1d_2}$ is always positive.
	Note that for the particular case $c = 0$ we have that in (\ref{eq: Je}) the error signals $e_1$ and $e_2$ do not depend on $e_3$ and the points $e_1 = e_2 = 0$ are stable regardless of $e_3$. In fact, in such a case we have a \emph{flexible shape} where the steady state of $e_3$ will depend on the initial conditions $p(0)$, i.e., the agents converge to the set (\ref{eq: noshape}).
\begin{remark}
	Note that by just adding a bias in a range sensor, no matter how small it is, we are linking the dynamics of the error signals $e_1$ and $e_2$ with $e_3$. This has an important implication since in practice it is not common (or arguably possible) to have perfect measurements in the agents' sensors. Therefore, to have a \emph{truly} flexible formation in practice would be quite challenging without any further action in (\ref{eq: 3flex}). A technique employing estimators in order to remove the biases has been introduced in \cite{MarCaoJa15}.
\end{remark}
In order to check the stability of the origin of system (\ref{eq: eaug}) we compute the characteristic polynomial of (\ref{eq: Je})
\begin{align}
	P(\lambda) &= -\lambda(2+ca+\lambda)^2 - 2ca(1-ca) - 2ca(2+ca+\lambda) \nonumber \\ &+\lambda(1-ca)^2 \nonumber \\
	&=(\lambda + 1)(\lambda + 3)(\lambda + 2ca), \label{eq: Pl}
\end{align}
therefore we can conclude that for any positive (negative) $c$ the equilibrium $\mathcal{U}_d$ is locally exponentially stable (unstable). The same analysis
can be done for checking the stability of $\mathcal{U}_u$ which is unstable (stable) for positive (negative) values of $c$. We summarize these findings with the following proposition.
\begin{prop}
	Consider system (\ref{eq: 3flexmu}) for positive values of $c=\mu_1=-\mu_2$, then the equilibria $\mathcal{U}_d$ and $\mathcal{U}_u$ for system (\ref{eq: ezdynm}) are (exponentially) stable and unstable respectively. The converse holds for negative values of $c$.
	\label{pr: 1}
\end{prop}
\begin{remark}
	For negative values of $c$ we have that the agents not only converge to a collinear formation but by inspecting (\ref{eq: 3flexmu}) we have that $\dot p_i(t) \to z^*\frac{3c}{2},$ as $t\to\infty, \forall i\in\{1,2,3\}$, i.e., the agents will travel with a constant velocity.
\end{remark}

\begin{remark}
	Checking the eigenvalues given by $P(\lambda)$ show that one cannot increase the convergence speed by increasing $c$. Conversely it can be made arbitrarily slow by choosing a small positive $c$. In fact, the effect of a small bias in the range sensors of system (\ref{eq: 3flex}) will be practically noticed after a sufficiently long time. 
\end{remark}

\section{Flexible triangular shape}
\label{sec: trian}
In this section we are going to show that the previous setup for collinear formations is just a particular case of a general setup where we can design controllers to achieve a prescribed angle between $z_1$ and $z_2$.

Consider the following system
\begin{equation}
	\begin{cases}
	\dot p_1 &= -\hat z_1 e_1 \\
	\dot p_2 &= \hat z_1 e_1 -\hat z_2 e_2 + W(\theta)\hat z_1 - \hat z_2\\
	\dot p_3 &= \hat z_2 e_2.
	\end{cases}
	\label{eq: 3flexW},
\end{equation}
where 
\begin{equation}
	W(\theta) = \begin{bmatrix}\cos\theta & -\sin\theta \\
\sin\theta & \cos\theta \end{bmatrix},
\end{equation}
is a rotation matrix. One can check that the two cases $\mu_1 = -\mu_2 = 1$ and $\mu_1 = \mu_2 = -1$ in (\ref{eq: 3flexmu}) are in fact particular cases of (\ref{eq: 3flexW}) with $\theta = 0$ and $\pm\pi$ radians respectively. In fact, the angle $\theta$ defines the prescribed angle between $\hat z_1$ and $\hat z_2$ in a target formation and note that the agents are not required to measure the actual angle between $z_1$ and $z_2$. We proceed to show that the agents indeed can employ their own local frames.
\begin{lem}
	The agents in system (\ref{eq: 3flexW}) can employ their own local frames of coordinates.
\end{lem}
\begin{pf}
	Consider $y_{ij}$ for the position of agent $j$ with respect to a fixed local frame of coordinates for agent $i$ with $j,i\in\{1, 2, 3\}$. Therefore, there exists a rotation matrix $R_i$ and a translation vector $\tau_i$ such that $y_{ij} = R_ip_j + \tau_i$. Note that the vectors $\tau_i$ with $i\in\{1,2,3\}$ vanish for the calculation of $z_1$ and $z_2$ and the rotation matrices do not affect the value of $e_1$ and $e_2$. Hence, agent $2$ measures the signals $R_2z_1$, $R_2z_2$ and we have that $W(\theta)R_2z_2 = R_2W(\theta)z_2$ since rotation matrices commute in 2D. Thus, for agent $2$ in global coordinates we have that
\begin{align}
\dot p_2 &= R_2^T \left(R_2\hat z_1 e_1 -R_2\hat z_2 e_2 + R_2W(\theta)\hat z_1 - R_2\hat z_2 \right) \nonumber \\
&= \hat z_1 e_1 -\hat z_2 e_2 + W(\theta)\hat z_1 - \hat z_2.
\end{align}
	Therefore, one can conclude that the control action of agent $2$ does not need any global information, e.g., a common frame of coordinates. The same reasoning can be applied to agents $1$ and $3$. $\blacksquare$.
\end{pf}

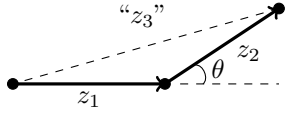
\begin{figure}
	\centering
\begin{tikzpicture}[line join=round]
\filldraw(0,0) circle (2pt);
\filldraw(2,0) circle (2pt);
\filldraw(3.5,1) circle (2pt);
\draw[very thick,arrows=->](0,0)--(2,0);
\draw[very thick,arrows=->](2,0)--(3.5,1);
\draw[dashed](3.5,1)--(0,0);
\draw[dashed](2.2,0)--(3.5,0);
\node at (1,-.2) {$z_1$};\node at (3.1,.4) {$z_2$};\node at (1.7,.9) {``$z_3$''};\draw[-] (2.5,0) to [bend left=-75] (2.4,.25);\node at (2.7,.2) {$\theta$};\end{tikzpicture}
	\caption{The triangular shape can be defined by setting the norms $||z_1|| = d_1$, $||z_2||=d_2$ and the angle $\theta$ between $z_1$ and $z_2$.} 
	\label{fig: trian}
\end{figure}

For the sake of convenience for the stability analysis, let us reformulate (\ref{eq: 3flexW}) as
\begin{equation}
	\begin{cases}
	\dot p_1 &= -\hat z_1 e_1 \\
		\dot p_2 &= \hat z_1 e_1 -\hat z_2 e_2 + W(\frac{\theta}{2})\hat z_1 - W(\frac{\theta}{2})^T\hat z_2\\
	\dot p_3 &= \hat z_2 e_2,
	\end{cases}
	\label{eq: 3flexW2}.
\end{equation}
Note that we are still defining the same desired steady-state angle between $\hat z_1$ and $\hat z_2$. The corresponding \emph{augmented} error system derived from (\ref{eq: 3flexW2}), with\footnote{Note $z_1$ points at agent $2$ and $z_2$ starts at agent $2$, therefore we will have $\cos(\pi - \theta) = -\cos(\theta)$ in the law of cosines for Figure \ref{fig: trian}.} $d_3^2 = d_1^2+d_2^2+2d_1d_2\cos\theta$, is given by
\begin{equation}
	\begin{cases}
		\dot e_1 &= -2e_1 + \hat z_1^T\hat z_2e_2 - \hat z_1^TW(\frac{\theta}{2})\hat z_1 + \hat z_1^TW(\frac{\theta}{2})^T \hat z_2 \\
		\dot e_2 &= -2e_2 + \hat z_2^T\hat z_1e_1 - \hat z_2^TW(\frac{\theta}{2})^T\hat z_2 + \hat z_2^TW(\frac{\theta}{2}) \hat z_1 \\
\dot e_3 &= \hat z_3^T\hat z_2e_2 + \hat z_3^T\hat z_1e_1.
	\end{cases}
	\label{eq: edynmW}
\end{equation}

In order to analyze the stability of the linearization of (\ref{eq: edynmW}) around $e = 0$ we need to work out some technical results first.
\begin{lem}
Consider the matrix
	\begin{equation}
	M = \begin{bmatrix}
	p_1 & p_2 & a \\
	p_2 & p_1 & a \\
	b & c & 0
	\end{bmatrix},
	\nonumber
	\end{equation}
	where $p_1, p_2, a, b, c\in\R$ with $p_1 > 0$, $p_1 > p_2$ and $p_1^2 > p_2^2$, so the second principal minor of $M$ is positive definite. Consider a small perturbation $a < 0$. If $(b + c) > 0$, then $-M$ is Hurwitz.
	\label{lem: 1}
\end{lem}
\begin{pf}
	First we note that if $a = 0$ the three eigenvalues of $M$ are defined by the two positive eigenvalues $\lambda_1$ and $\lambda_2$ of its 2x2 leading principal submatrix and $\lambda_3 = 0$. Therefore a small perturbation $a$ will not change the sign of the positive eigenvalues of $M$ but we do not know about the sensitivity of the zero eigenvalue $\lambda_3$ with respect to the disturbance $a$. In order to check such sensitivity we derive the characteristic polynomial of $M$
\begin{align}
	P(\lambda) &= -\lambda(p_1-\lambda)^2 + abp_2+acp_2-ab(p_1-\lambda)+\lambda p_2^2 \nonumber \\ &-ac(p_1-\lambda) \nonumber \\
	&= \lambda \left(p_2^2 - (p_1-\lambda)^2 + a(b+c)\right) + a(b+c)(p_2-p_1). \label{eq: poly}
\end{align}
	Note that $a$ can be chosen arbitrarily small. Since $\lambda_3$ is a continuous function of $a$, it can also be made arbitrarily small, so $|p_1| \gg |\lambda_3|$ and $|(p_2^2 - p_1^2)| \gg |a(b+c)|$. Then, without affecting the analysis of the sign of the perturbed $\tilde\lambda_3$, we can approximate the first bracket in (\ref{eq: poly}) by $(p_2^2 - p_1^2)$. Therefore we can look at 
\begin{align}
	P(\tilde\lambda_3) \approx \tilde\lambda_3 (p_2^2 - p_1^2) + a(b+c)(p_2-p_1) = 0,
\end{align}
in order to check the sign of the perturbed $\tilde\lambda_3$. We note that under the assumptions in the statement of the lemma one concludes that for a small negative $a$ the eigenvalue $\tilde\lambda_3$ becomes positive. Hence $-M$ is Hurwitz.$\blacksquare$
\end{pf}
\begin{remark}
	The Jacobian in (\ref{eq: Je}) fits the description of the matrix $M$ in Lemma \ref{lem: 1}. Note that from Proposition \ref{pr: 1} the sign of $\tilde\lambda_3$ from (\ref{eq: Je}) only depends on the sign of $a$ and not on its magnitude.
\end{remark}

We now work out a series of calculations that will be needed for the calculation of the Jacobian matrix of (\ref{eq: edynmW}) at $e=0$.
Consider the expression
\begin{equation}
	\hat z_2^TW(\alpha)\hat z_1 = \cos(\gamma - \alpha),
	\label{eq: foo}
\end{equation}
where $\gamma$ is the actual angle between $z_1$ and $z_2$ and $\alpha\in\R$ is a constant angle.
We also have that
\begin{equation}
	\frac{\partial \hat z_2^TW(\alpha)\hat z_1}{\partial e} = \frac{\partial \cos(\gamma(e) - \alpha)}{\partial e},
	\label{eq: z2z1e}
\end{equation}
with
\begin{equation}
	\gamma(e) = \arccos\left(\frac{(e_3+d_3)^2-(e_1+d_1)^2-(e_2+d_2)^2}{2(e_1+d_1)(e_2+d_2)}\right),
\end{equation}
and note that $\gamma(0) = \theta$. Therefore by applying the chain rule
\begin{align}
	&\frac{\partial \cos(\gamma(e) - \alpha)}{\partial e_1}|_{e=0} 
	= -\sin(\gamma(e)-\alpha)\,\frac{\partial\gamma(e)}{\partial e_1}|_{e=0} \nonumber \\
	&= \sin(\theta-\alpha) \frac{1}{\sqrt{1-\cos^2\theta}}\, \left(-\frac{2d_1}{2d_1d_2}-\frac{d_3^2-d_1^2-d_2^2}{d_1^2d_2}\right) \nonumber \\
	&= -\frac{\sin(\theta-\alpha)}{\sin(\theta)} \left(\frac{d_1+d_2\cos\theta}{d_1d_2}\right) =: a_1
	\label{eq: a1}
\end{align}
\begin{align}
	\frac{\partial \cos(\gamma(e) - \alpha)}{\partial e_2}|_{e=0} 
	&= -\frac{\sin(\theta-\alpha)}{\sin(\theta)} \left(\frac{d_2+d_1\cos\theta}{d_1d_2}\right) =: a_2
	\label{eq: a2}
\end{align}
\begin{align}
	\frac{\partial \cos(\gamma(e) - \alpha)}{\partial e_3}|_{e=0} &=
	\frac{\sin(\theta - \alpha)}{\sin{\theta}}\,\frac{d_3}{d_1d_2} =: a_3.
	\label{eq: a3a}
\end{align}

\begin{remark}
We highlight again that $\alpha$ is a constant parameter that can be chosen arbitrarily. For example, the computation of $a_3$ for $\theta = 0$, i.e., $d_3 = d_1 + d_2$, and $\alpha = 0$ becomes
\begin{equation}
	 \lim_{\theta\to 0}a_3(\theta) = \lim_{\theta\to 0} \frac{\sin(\theta - 0)}{\sin{\theta}}\,\frac{d_1+d_2}{d_1d_2} = \frac{d_1+d_2}{d_1d_2}, \label{eq: a3lim}
	\end{equation}
	which is consistent with the values obtained for the Jacobian in (\ref{eq: Je}).
\end{remark}

Similarly as in (\ref{eq: 3flexmu}) with (\ref{eq: muc}), we also introduce a gain $c > 0$ in (\ref{eq: 3flexW2}) multiplying the terms $W(\frac{\theta}{2})\hat z_1$ and $W(\frac{\theta}{2})^T\hat z_2$. Now we are ready to present our main result.
\begin{thm}
For the system
	\begin{equation}
	\begin{cases}
	\dot p_1 &= -\hat z_1 e_1 \\
		\dot p_2 &= \hat z_1 e_1 -\hat z_2 e_2 + c\left(W(\frac{\theta}{2})\hat z_1 - W(\frac{\theta}{2})^T\hat z_2\right)\\
	\dot p_3 &= \hat z_2 e_2,
	\end{cases}
	\label{eq: 3flexW3}
\end{equation}
its equilibrium defined by the set $\mathcal{U} \dfb \{p_1, p_2, p_3 : W(\theta)\hat z_1 - \hat z_2 = 0, ||z_1|| = d_1, ||z_2||=d_2\}$, for every $\theta\in(-\pi, \pi)$ is locally exponentially stable for a sufficiently small $c > 0$. 
	\label{th: 1}
\end{thm}
\begin{pf}
	We first note that the study of the stability of the equilibrium $\mathcal{U}_d$ corresponds to checking the eigenvalues of the Jacobian of the autonomous \emph{augmented} error system (\ref{eq: edynmW}), with the introduced gain $c$, at $e = 0$. The Jacobian is given by
\begin{equation}
	J_e = \begin{bmatrix}
-2+ca_1 & \cos(\theta)+ca_2 & ca_3 \\
\cos(\theta)+ca_1 & -2+ca_2 & ca_3 \\
		-\frac{d_1 + d_2\cos\theta}{d_3} & -\frac{d_2 + d_1\cos\theta}{d_3} & 0
	\end{bmatrix},\label{eq: Je2}
\end{equation}
where $a_1, a_2$ and $a_3$ are as in (\ref{eq: a1})-(\ref{eq: a3a}) respectively.
We notice that the following expression
\begin{align}
	&- (\frac{d_1 + d_2\cos\theta}{d_3} + \frac{d_2 + d_1\cos\theta}{d_3}) = \nonumber \\ &
	- \frac{d_1 (1 + \cos\theta) + d_2(1+\cos\theta)}{d_3},
\end{align}
	is always negative for $\theta \in (-\pi, \pi)$. Since $\alpha = \frac{\theta}{2}$ one can also check that (\ref{eq: a3a}) is always positive. Indeed, the second principal minor is not symmetric (negative definite) anymore because of the added small perturbations $ca_1$ and $ca_2$, but its eigenvalues are still with real negative part and bounded from zero for all $\theta$. Furthermore, the diagonal elements and off-diagonal elements of the second principal minor are very similar respectively for a very small $c > 0$. Therefore from the eigenvalue sensitivity analysis in Lemma \ref{lem: 1} one can derive that indeed (\ref{eq: Je2}) is Hurwitz for a very small gain $c>0$. Note that the local exponential convergence of the augmented $e(t)$ to the origin implies that the agents' velocities in (\ref{eq: 3flexW2}) converge to zero as well, so the agents converge to fixed positions.$\blacksquare$
\end{pf}
\begin{remark}
Note that in system (\ref{eq: 3flexW3}), the set of collinear positions for all the agents is not invariant. Therefore, it outperforms in that sense the distance-based control of rigid formations, where if the agents start collinear, they remain collinear.
\end{remark}

\section{Simulations}
\label{sec: sim}
\begin{figure}
\centering
\includegraphics[width=1\columnwidth]{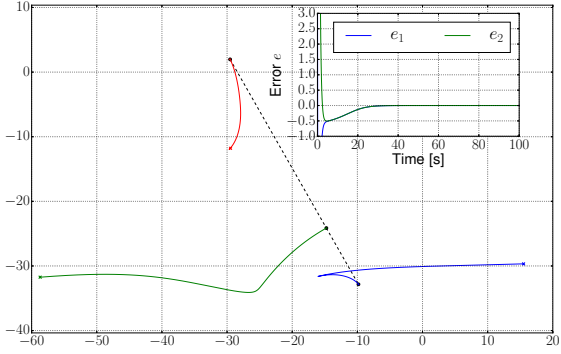}
	\caption{Simulation result from Proposition \ref{pr: 1}. If $\mu_1 = -\mu_2 = 1$, then the equilibrium set $\mathcal{U}_d$ is stable and all the agents stop eventually once the distance errors converge to zero (shown in the inner plot). In this equilibrium, we recall that $\hat z_1 = \hat z_2$, therefore we have that the agent $2$ marked in green will finish between the other two agents.}
\label{fig: ud}
\end{figure}

\begin{figure}
\centering
\includegraphics[width=1\columnwidth]{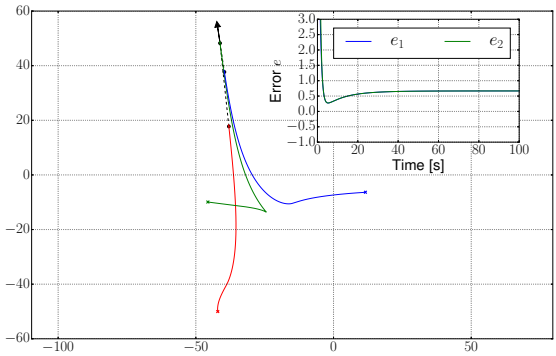}
	\caption{Simulation result from Proposition \ref{pr: 1}. If $\mu_2 = -\mu_1 = 1$, then the equilibrium set $\mathcal{U}_u$ is stable and the formation describes a steady-state collective motion (denoted by the black arrow) in the direction of $\hat z_2$. Note that $\hat z_1 = -\hat z_2$, therefore the agent $2$ marked in green is leading the motion. Both distance errors $e_1$ and $e_2$ (in the inner plot) converge to $\frac{2}{3}$, regardless of the prescribed distances.}
\label{fig: uu}
\end{figure}

In this section we validate the results in Proposition \ref{pr: 1} and Theorem \ref{th: 1}. We first show in Figure \ref{fig: ud} that for $\mu_1 = -\mu_2 = 1$ in system (\ref{eq: 3flexmu}) the equilibrium $\mathcal{U}_d$ is stable, where all the agents converge to collinear positions with $\hat z_1 = \hat z_2$. The desired distances are $d_1 = 30$ and $d_2 = 10$, and the agents $1, 2$ and $3$ are marked with red, green and blue colors respectively. We then show in Figure \ref{fig: uu} that the equilibrium $\mathcal{U}_u$ is stable for $\mu_2 = -\mu_1 = 1$. Note that the agents do not stop moving and the distance errors converge both to $\frac{2}{3}$.

\begin{figure}
\centering
\includegraphics[width=1\columnwidth]{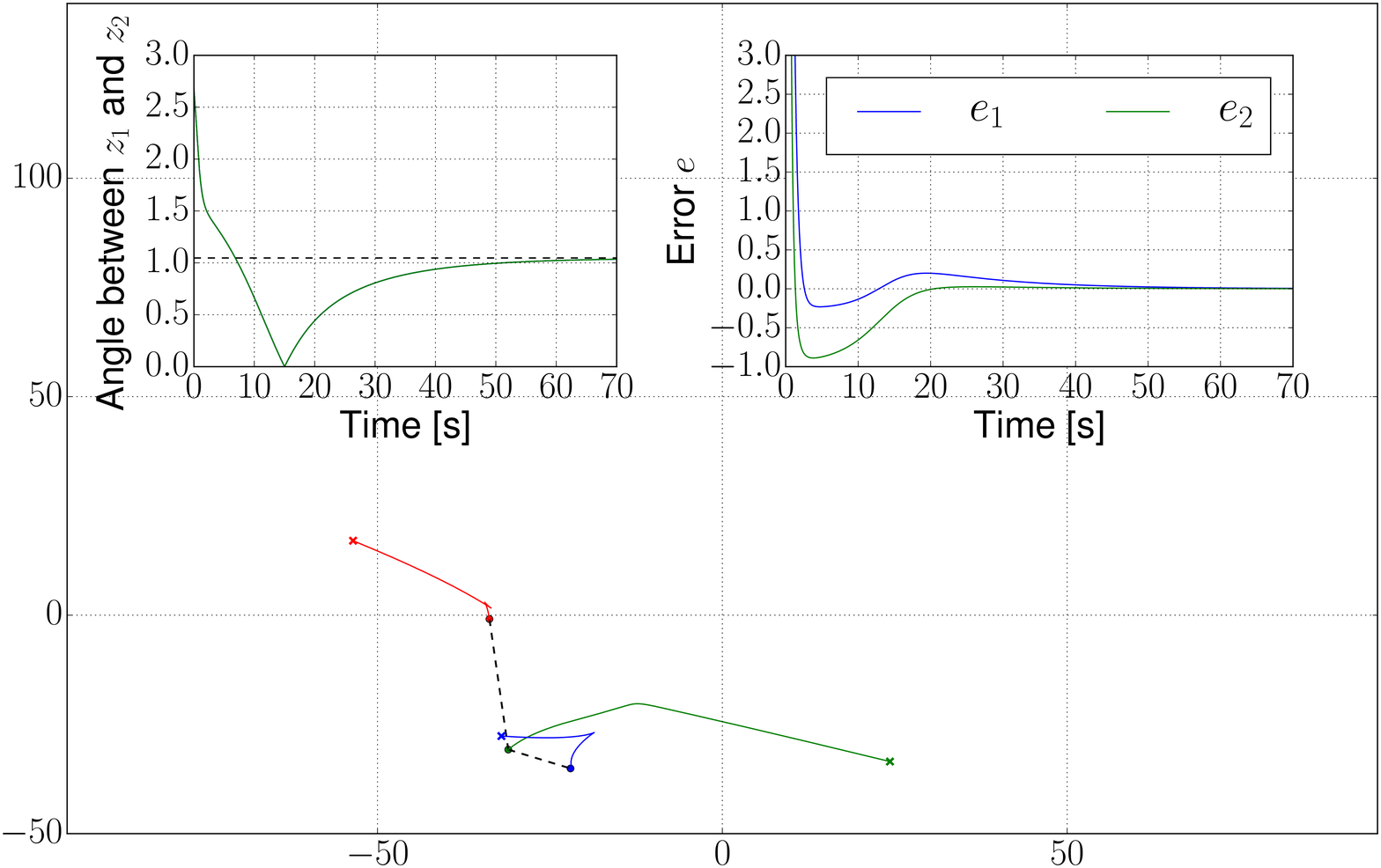}
	\caption{Simulation result from Theorem \ref{th: 1}. In addition to the distance control between agents $1,2$ and $2,3$ (right inner plot), the agent $2$ marked in green color is controlling the angle between $\hat z_1$ and $\hat z_2$ (left inner plot). Note that the final orientation of the shape is not controlled and depends on the initial conditions.}
\label{fig: tri}
\end{figure}

We move on to control a triangular shape by employing the results from Theorem \ref{th: 1}. The desired angle betwen $\hat z_1$ and $\hat z_2$ has been set to $60$ degrees. The evolution of the agents can be seen in Figure \ref{fig: tri}, where the left inner plot shows the evolution of the \emph{inter-angle} and the right inner plot the evolution of the distance errors.

\section{Conclusions}
\label{sec: con}
We have presented a formation control algorithm for achieving triangular shapes. The algorithm is based on a new technique derived from the biased distance-based gradient descent control of two links. The algorithm enjoys the advantages of both position-based and distance-based formation controls. Current research is aimed at providing a systematic approach for achieving stable arbitrary shapes consisting of more than three agents in non-rigid formation. We have already made some progress for chain topologies \cite{ifacnew}.
\bibliography{nonrigid}             

\begin{thebibliography}{14}
\providecommand{\natexlab}[1]{#1}
\providecommand{\url}[1]{\texttt{#1}}
\providecommand{\urlprefix}{URL }
\expandafter\ifx\csname urlstyle\endcsname\relax
  \providecommand{\doi}[1]{doi:\discretionary{}{}{}#1}\else
  \providecommand{\doi}{doi:\discretionary{}{}{}\begingroup
  \urlstyle{rm}\Url}\fi

\bibitem[{Anderson et~al.(2008)Anderson, Yu, Fidan, and Hendrickx}]{AnYuFiHe08}
Anderson, B.D.O., Yu, C., Fidan, B., and Hendrickx, J. (2008).
\newblock Rigid graph control architectures for autonomous formations.
\newblock \emph{IEEE Control Systems Magazine}, 28, 48--63.

\bibitem[{Anderson et~al.(2007)Anderson, Yu, Dasgupta, and
  Morse}]{anderson2007control}
Anderson, B.D.O., Yu, C., Dasgupta, S., and Morse, A.S. (2007).
\newblock Control of a three-coleader formation in the plane.
\newblock \emph{Systems \& Control Letters}, 56(9), 573--578.

\bibitem[{Cao et~al.(2007)Cao, Morse, Yu, Anderson, and
  Dasgupta}]{cao2007controlling}
Cao, M., Morse, A.S., Yu, C., Anderson, B.D.O., and Dasgupta, S. (2007).
\newblock Controlling a triangular formation of mobile autonomous agents.
\newblock In \emph{Decision and Control, 2007 46th IEEE Conference on},
  3603--3608. IEEE.

\bibitem[{Cao et~al.(2008)Cao, Yu, Morse, Anderson, and
  Dasgupta}]{cao2008generalized}
Cao, M., Yu, C., Morse, A.S., Anderson, B.D.O., and Dasgupta, S. (2008).
\newblock Generalized controller for directed triangle formations.
\newblock \emph{IFAC Proceedings Volumes}, 41(2), 6590--6595.

\bibitem[{Dimarogonas and Johansson(2008)}]{dimarogonas2008stability}
Dimarogonas, D.V. and Johansson, K.H. (2008).
\newblock On the stability of distance-based formation control.
\newblock In \emph{Decision and Control, 2008. CDC 2008. 47th IEEE Conference
  on}, 1200--1205. IEEE.

\bibitem[{Garcia~de Marina et~al.(2015)Garcia~de Marina, Cao, and
  Jayawardhana}]{MarCaoJa15}
Garcia~de Marina, H., Cao, M., and Jayawardhana, B. (2015).
\newblock Controlling rigid formations of mobile agents under inconsistent
  measurements.
\newblock \emph{Robotics, IEEE Transactions on}, 31(1), 31--39.

\bibitem[{Garcia~de Marina et~al.(2016)Garcia~de Marina, Jayawardhana, and
  Cao}]{Hector2016maneuvering}
Garcia~de Marina, H., Jayawardhana, B., and Cao, M. (2016).
\newblock Distributed rotational and translational maneuvering of rigid
  formations and their applications.
\newblock \emph{IEEE Transactions on Robotics}, 32(3), 684--697.

\bibitem[{Garcia~de Marina et~al.(2017)Garcia~de Marina, Jayawardhana, and
  Cao}]{ifacnew}
Garcia~de Marina, H., Jayawardhana, B., and Cao, M. (2017).
\newblock Distributed algorithm for controlling scale-free polygonal
  formations.
\newblock In \emph{proceedings of the 2017 IFAC World Congress}. IFAC.

\bibitem[{Krick et~al.(2009)Krick, Broucke, and Francis}]{KrBrFr08}
Krick, L., Broucke, M.E., and Francis, B.A. (2009).
\newblock Stabilization of infinitesimally rigid formations of multi-robot
  networks.
\newblock \emph{International Journal of Control}, 82, 423--439.

\bibitem[{Liu et~al.(2014)Liu, Garcia~de Marina, and Cao}]{liu2014controlling}
Liu, H., Garcia~de Marina, H., and Cao, M. (2014).
\newblock Controlling triangular formations of autonomous agents in finite time
  using coarse measurements.
\newblock In \emph{2014 IEEE International Conference on Robotics and
  Automation (ICRA)}, 3601--3606. IEEE.

\bibitem[{Mou et~al.(2016)Mou, Belabbas, Morse, Sun, and
  Anderson}]{SMA16TACsub}
Mou, S., Belabbas, M.A., Morse, A.S., Sun, Z., and Anderson, B.D.O. (2016).
\newblock Undirected rigid formations are problematic.
\newblock \emph{IEEE Transactions on Automatic Control}, 61(10), 2821--2836.

\bibitem[{Mou et~al.(2014)Mou, Morse, Belabbas, and Anderson}]{7039453}
Mou, S., Morse, A.S., Belabbas, M.A., and Anderson, B.D.O. (2014).
\newblock Undirected rigid formations are problematic.
\newblock In \emph{53rd IEEE Conference on Decision and Control}, 637--642.

\bibitem[{Oh et~al.(2015)Oh, Park, and Ahn}]{oh2015survey}
Oh, K.K., Park, M.C., and Ahn, H.S. (2015).
\newblock A survey of multi-agent formation control.
\newblock \emph{Automatica}, 53, 424--440.

\bibitem[{Sun et~al.(2016)Sun, Mou, Anderson, and Cao}]{sun2016exponential}
Sun, Z., Mou, S., Anderson, B.D.O., and Cao, M. (2016).
\newblock Exponential stability for formation control systems with generalized
  controllers: A unified approach.
\newblock \emph{Systems \& Control Letters}, 93, 50--57.

\end{thebibliography}

\end{document}